# A First Principles Electronic Structure Study of Quantum Size Effects in (111) Films of δ-Plutonium

A. K. Ray*[1] and J. C. Boettger[2]

[1]*Physics Department, The University of Texas at Arlington, Arlington, Texas 76019*
[2]*Applied Physics Division, Los Alamos National Laboratory, Los Alamos, New Mexico 87545*

First principles linear combinations of Gaussian type orbitals – fitting function (LCGTO-FF) electronic structure calculations are used to study thickness dependencies in the surface energies and work functions of ultra-thin (111) films of fcc δ-Pu, up to five layers thick. The calculations are carried out at both the scalar- and fully-relativistic (with and without spin-orbit coupling) levels of approximation. The surface energy is shown to be rapidly convergent, while the work function exhibits a strong quantum size effect for all thicknesses considered. The surface energy and work function of the semi-infinite solid are predicted to be 1.12 J/m$^2$ and 2.85 ± 0.20 eV, respectively, for the fully-relativistic case. These results are in substantial disagreement with results from previous electronic structure calculations. The present predictions are in fair agreement with the most recent experimental data for polycrystalline δ-Pu, namely 0.91 J/m$^2$ and 3.1-3.3 eV, for the surface energy and work function, respectively.

## 1. Introduction

During the past two decades, considerable theoretical efforts have been devoted to studying the electronic and geometric structures and related properties of surfaces to high accuracy. One of the many motivations for this burgeoning effort has been a desire to understand the detailed mechanisms that lead to surface corrosion in the presence of environmental gases; a problem that is not only scientifically and technologically challenging but also environmentally important. Such efforts are particularly important

akr@uta.edu

for systems like the actinides for which experimental work is relatively difficult to perform due to material problems and toxicity.

Among the actinides, plutonium metal exhibits properties that are unique.[1-6] First, Pu has, at least, six stable allotropes between room temperature and melting at atmospheric pressure, indicating that the valence electrons can hybridize into a number of complex bonding arrangements. Second, plutonium represents the boundary between the light actinides, Th to Pu, characterized by itinerant 5f electron behavior, and the heavy actinides, Am and beyond, characterized by localized 5f electron behavior. In fact, the high temperature fcc δ-phase of plutonium exhibits properties that are intermediate between the properties expected for the light and heavy actinides. These unusual aspects of the bonding in bulk Pu are apt to be enhanced at a surface or in a thin layer of Pu adsorbed on a substrate, as a result of the reduced atomic coordination of a surface atom and the narrow bandwidth of surface states. For this reason, Pu surfaces and films may provide a valuable source of information about the bonding in Pu. For studies like these, it is common practice to model the surface of a semi-infinite solid with an ultra-thin film (UTF) which is thin enough to be treated with high-precision density functional calculations, but is thick enough to model the intended surface realistically. Determination of an appropriate UTF thickness is complicated by the existence of quantum oscillations in UTF properties as a function of thickness; the so-called quantum size effect (QSE). These oscillations were first predicted by calculations on jellium films[7,8] and were subsequently confirmed by band structure calculations on free-standing UTFs composed of discrete atoms.[9-12] The adequacy of the UTF approximation will obviously depend on the size of any QSE in the relevant properties of the model film. Thus it is important to determine the magnitude of the QSE in a given UTF prior to using that UTF as a model for the surface.[13] This is particularly important for Pu films, since the strength of the QSE is expected to increase with the number of valence electrons.[7]

Although the ambient α-phase of Pu is known to be monoclinic with sixteen atoms per unit cell, there are a number of reasons for instead focussing on UTFs that are extracted from the high-temperature fcc δ-phase of Pu, which can be stabilized at room temperature through the addition of very small amounts of particular impurities.[4] First, as noted above, δ-Pu exhibits properties that are intermediate between the light- and heavy-actinides. In addition, grazing-incidence photoemission studies, combined with the calculations of Eriksson *et al.*[14], suggest the existence of a small-moment δ-like surface on α-Pu. Our work on Pu monolayers has also indicated the possibility of such a surface.[15] Recently, high-purity ultrathin layers of Pu deposited on Mg were studied by X-ray photoelectron (XPS) and high-resolution valence band (UPS) spectroscopy by Gouder *et al.*[16] They found that the degree of delocalization of the 5f states depends in a very dramatic way on the layer thickness and that the itinerant character of the 5f states is gradually lost with reduced thickness, suggesting that the thinner films are δ-like. Finally, it may be possible to study 5f localization in Pu layers through adsorption on a series of carefully selected substrates; in which case, the adsorbed layers are more likely to be δ-like than α-like. Hence, in this study, the properties of hexagonal Pu n-layers (n=1-5) corresponding to the (111) surface of δ-Pu have been determined with a *fully relativistic* version of the linear combinations of Gaussian type orbitals fitting function (LCGTO-FF) method, as embodied in the program GTOFF.[17]

**2. Computational Method**

The LCGTO-FF technique is an all-electron, full-potential electronic structure method which is characterized by the use of two auxiliary GTO basis sets to expand the charge density and the exchange-correlation (XC) integral kernels, here using the Perdew-Wang generalized-gradient-approximation (GGA) model[18] to density functional theory (DFT)[19]. The charge fitting function coefficients are determined variationally, by minimizing the error in the Coulomb energy, while the XC coefficients are obtained from a constrained least squares fit. In its non-relativistic form, the LCGTO-FF method is known to yield results that are comparable to results produced by other all-electron, full-potential DFT methods. Unlike many other electronic structure methods, however, the LCGTO-FF method is directly applicable to such diverse systems as isolated clusters of atoms,[20] polymer chains,[21] free standing ultra-thin films[22] and crystalline solids.[17]

The relativistic implementation of the LCGTO-FF method has progressed through several stages over the years. Scalar-relativity was initially implemented in GTOFF[23] using a nuclear-only Douglas-Kroll-Hess transformation[24] that neglected all terms involving cross-products of the momentum operator. The implementation of relativity in

GTOFF was subsequently extended to include all scalar-relativistic cross-product terms and spin-orbit coupling terms produced by the nuclear-only DKH transformation. Although this nuclear-only DKH approximation was shown to produce reliable scalar-relativistic results for atoms, molecules and solids,[25] it consistently overestimated the effects of spin-orbit coupling, effectively providing an upper bound on such effects.[26] This limitation was overcome via the development of the screened-nuclear-spin-orbit (SNSO) approximation,[26] which approximately incorporates two-electron spin-orbit coupling effects, without increasing the computational effort relative to the nuclear-only approximation. For atoms ranging from Ce to Pu, the SNSO approximation yields spin-orbit splittings that agree with results from a numerical solution of the Dirac-Kohn-Sham (DKS) equation to within about six percent, the difference with the DKS results ranging from 0.1 per cent for the 4d orbital of the Ce atom to 6.1 per cent for the 6p orbital of the Pu atom.[26] For the fcc phases of the light actinides, Th-Pu, the SNSO approximation also yields atomic volumes and bulk moduli that are in excellent agreement with FLAPW results.[26] We have also applied this method to investigate the electronic structure properties of three fluorite-structure actinide oxides, namely $ThO_2$, $UO_2$, and $PuO_2$, and their clean and hydroxylated (111) surfaces.[27]

The overall precision of any LCGTO-FF calculation is, to a large extent, determined by the selection of the three basis sets. In this work, the orbital basis set used for Pu started with a 23s20p15d11f uncontracted basis set derived from an atomic basis set.[28] This basis set was contracted into a "bulk" 17s14p11d7f basis with coefficients taken from scalar-relativistic atomic calculations using the GGA model. A richer "surface" basis set was then obtained by augmenting the "bulk" basis with an s-type function with an exponent of 0.07, a $p_z$-type function with an exponent of 0.08, and a d-type function with an exponent of 0.12. The "surface" basis set was used for the monolayer, the dilayer, and the surface layers of all thicker films. The "bulk" basis was used for the interior layers of the 3-, 4-, and 5-layer films. The "bulk" charge and XC basis sets were 25s2d and 21s2d, respectively. The "surface" charge and XC basis sets were each augmented with two $p_z$-type functions.[29]

Although the existence of magnetic moments in Pu metal is a subject of great controversy at this time,[30] traditional DFT calculations clearly predict spin-polarization in δ-Pu[31] and it seems likely that magnetic moments play an important role in the unusual properties exhibited by bulk Pu. The picture is less clear for the surface of δ-Pu, partly because of the very limited studies in the literature. Also, unfortunately, as films become thicker, the complexity of the magnetic ordering increases, making such calculations prohibitive. Hence, the effects of collinear, ferromagnetic spin-polarization have been studied only for the Pu mono- and dilayers. (It also should be noted that the current version of GTOFF does not allow a determination of orbital moments or a decomposition of the electron density into angular momentum components.) Because of severe demands on computational resources, the lattice constant was not optimized and all computations have been carried out at the experimental lattice constant of bulk δ - Pu. The two-dimensional Brillouin zone was sampled on a uniform mesh with 19 irreducible points. For each calculation, the SCF cycle was iterated until the total energy was stable to within 0.01 mRy.

## 3. Results

For each film, total energies, cohesive energies, surface energies, and work functions have been calculated, both with and without spin-orbit coupling included, *i.e.* at the scalar- and fully-relativistic levels. For the monolayer and the dilayer, we also carried out spin-polarized calculations. The spin-polarization energy for the monolayer is 2.39 eV at the scalar-relativistic level and 1.30 eV at the fully-relativistic level. The spin magnetic moments of the spin-polarized monolayers are predicted to be 6.69 $\mu_B$ and 6.35 $\mu_B$ at the scalar- and fully-relativisitic levels, respectively. In our previous work,[15] the spin magnetic moment for a geometry-optimized Pu(111) monolayer (with a nearest-neighbor-distance of 4.89 a.u versus 6.1936 a.u. used here) was found to be about 1.85 $\mu_B$ at the scalar-relativistic level (note that simultaneous inclusion of spin-polarization and spin-orbit coupling was not implemented in GTOFF at that time). We also observed in that work that the spin magnetic moment increases monotonically with the lattice constant and thus, the larger spin moment found in this work is reasonable given the increased lattice constant. For the dilayer, the scalar- and fully-relativistic calculations yield spin-polarization energies of 3.45 eV and 1.63 eV, respectively, corresponding to spin magnetic moments of 5.86 $\mu_B$ and 5.23 $\mu_B$. Thus, the spin magnetic moment decreases by 0.83 $\mu_B$ in going from the monolayer to the dilayer at the scalar-relativistic level and by 1.12 $\mu_B$ at the fully-relativistic level.

The cohesive energies for the n-layers have been calculated here relative to two reference systems; (1) n monolayers, and (2) an (n-1)-layer plus a single monolayer (the incremental energy). Table 1 shows the results for the total energies and the cohesive energies. For the dilayer, of course, the cohesive energy can only be calculated with respect to two monolayers. We note that the dilayer is bound with respect to the monolayer at all levels of theory and that spin-polarization lowers the cohesive energy by fifty-five percent at the SR level and forty-two percent at the SO level. On the other hand, SO coupling increases the cohesive energy of the spin-polarized dilayer from 1.1 eV to 1.3 eV but reduces the cohesive energy of the non-spin-polarized dilayer from 2.42 eV to 2.26 eV. For all films, several features are evident: first, all cohesive energies are positive, indicating that all higher layers are bound relative to the monolayer; second, the cohesive energy for a Pu n-layer computed relative to the monolayer increases monotonically; and finally, the cohesive energy for an n-layer computed relative to an (n-1) layer and a monolayer shows an oscillatory pattern.

The surface energy may be estimated from an n-layer calculation as:[32]

$$E_s = (1/2)\,[E(n) - n\,E_B] \qquad (1)$$

where $E(n)$ is the total energy of the n-layer film and $E_B$ is the energy of the infinite crystal. If n is sufficiently large and $E(n)$ and $E_B$ are known to infinite precision, Eq. (1) is exact. If, however, the bulk and the film calculations are not entirely consistent with each other, $E_s$ will diverge linearly with increasing n.[33] Stable and internally consistent estimates of $E_s$ and $E_B$ can, however, be extracted from a series of values of $E(n)$ versus n via a linear least-squares fit to:[32]

$$E(n) = E_B n + 2E_s \qquad (2)$$

To obtain an optimal result, the fit to Eq. (2) should only be applied to films which include, at least, one bulk-like layer, *i.e.* n>2. We have independently applied this procedure to the NSP films at both the scalar- and fully-relativistic levels. We obtain $E_B$ = -29512.76623 a.u. and $E_s$ = 0.78 eV/atom = 1.34 J/m$^2$ for the scalar-relativistic calculations, and $E_B$ = -29616.79396 a.u. and $E_s$ = 0.65 eV/atom = 1.12 J/m$^2$ for the fully-relativistic calculations. Thus, the semi-infinite surface energy decreases by close to seventeen percent from the scalar-relativistic to the fully relativistic case. Durakiewicz[34] has recently used a semi-empirical model to estimate the surface energy of δ-Pu, obtaining $E_s$ = 0.91 J/m$^2$, in reasonable agreement with the present fully-relativistic result.

The surface energy for each layer has been computed using the calculated n-layer total energy and the appropriate fitted bulk energy. Results are listed in table 2 and are plotted in figure 1. Two features of the surface energies are evident from these results: (1) The surface energies obtained with the scalar- and fully-relativistic calculations show identical oscillatory patterns and (2) The 3 to 5-layer surface energies are nearly converged to the semi-infinite surface energy. *From these results, we infer that a 3-layer film may be sufficient for future atomic and molecular adsorption studies on Pu films, if the primary quantity of interest is the chemisorption energy.*

As mentioned in the introduction, the second primary quantity of interest in these calculations is the work function, calculated as the negative of the Fermi energy. The results obtained here are shown in table 2 and figure 2. Two qualitative trends are evident; (1) the work function shows an odd-even oscillatory pattern at both the NSP-SR and the NSP-SO levels and (2) the calculated work function exhibits a strong QSE over the full range of thicknesses considered here. The latter trend suggests that film thicknesses greater than n = 5 will be required for any chemisorption investigation that requires an accurate prediction of the adsorbate-induced work function shift. In spite of the strong work function QSE, it is clear that the work function for the semi-infinite solid should be roughly 2.85 ± 0.20 eV, at the NSP-SO level. Quite recently, Durakiewicz *et al.*,[35] measured the work function of δ-Pu for various degrees of surface oxidation. Based on their measurements, they obtained a preliminary estimate for the work function of a clean polycrystalline sample of δ-Pu of 3.1-3.3 eV, in reasonable agreement with our NSP-SO result for the (111) surface. Some caution is, however, appropriate since the Pu 5f electrons of the surface layers are likely to be more localized than those of the bulk and the interpretation of DFT results needs to be done properly.

The present values for the monolayer work functions, 3.08 eV and 2.99 eV, are notably smaller than our previous results for geometry-optimized Pu(111) monolayers, 4.74 eV and 4.28 eV,[15] due to large differences in the nearest-neighbor distances. Hao *et al.*[14] carried out film-linearized-muffin-tin-orbital (FLMTO) calculations for 5-layer films representing the (111) and (100) surfaces of δ–Pu. In those calculations, they treated the 6p electrons either as core electrons or valence electrons, obtaining work functions of 4.14 eV or 8.4 eV, respectively for the (111) film, significantly larger than the present NSP-SO value for the 5-layer work function (2.87 eV).

## 4. Conclusion

The linear combinations of Gaussian type orbitals - fitting function (LCGTO-FF) method has been used to study thickness dependencies in the properties of ultra-thin (111) films of fcc δ-Pu, up to five layers thick. At the fully-relativistic level, the surface energy and work function of the (111) surface are predicted to be 1.12 J/m$^2$ and 2.85 ± 0.20 eV, respectively. These predictions compare well with the most recent semi-empirical estimate for the surface energy (0.91 J/m$^2$)[34] and a preliminary measurement of the polycrystalline work function (3.1-3.3 eV).[35] The surface energy is rapidly convergent with respect to the film thickness, suggesting that a 3-layer may provide an adequate model for estimating chemisorption energies on the (111) surface of δ-Pu. On the other hand, the work function exhibits a significant quantum size effect for all of the films considered, indicating that a much thicker film would be required to realistically estimate adsorbate-induced work function shifts on that surface.

## Acknowledgments

We thank T. Durakiewicz for sharing his unpublished results (Refs. 34 and 35) with us. The work of AKR was supported by the Chemical Sciences, Geosciences, and Biosciences Division, Office of Basic Energy Sciences, Office of Science, U. S. Department of Energy (Grant No. DE-FG02-03ER15409) and the Welch Foundation, Houston, Texas (Grant No. Y-1525). The work of JCB was supported by the U. S. Department of Energy under contract W-7405-ENG-36 and the LDRD program at Los Alamos National Laboratory.

**Table 1**. Total energies and cohesive energies for Pu$_n$ (n = 1-5) layers. Total energies are in atomic units and cohesive energies, indicated in brackets, are in eV. The first number represents cohesive energies with respect to n monolayers, the second number, with respect to the monolayer and (n-1) layer.

| System | NSP-SR | SP-SR | NSP-SO | SP-SO |
|---|---|---|---|---|
| 1-layer | -29512.690243 | -29512.777988 | -29616.726443 | -29616.774180 |
| 2-layer | -59025.469338 | -59025.596318 | -59233.535959 | -59233.596054 |
|  | (2.42) | (1.10) | (2.26) | (1.30) |
| 3-layer | -88538.241698 |  | -88850.335199 |  |
|  | (4.65, 2.23) |  | (4.24, 1.98) |  |
| 4-layer | -118051.006651 |  | -118467.125547 |  |
|  | (6.68, 2.03) |  | (5.98, 1.74) |  |
| 5-layer | -147563.774150 |  | -148083.923104 |  |
|  | (8.78, 2.10) |  | (7.91, 1.41) |  |

**Table 2**. Surface energies and work functions in eV for Pu n-layers (n = 1-5).

| System | Surface Energy | | Work Function | |
| --- | --- | --- | --- | --- |
| | NSP-SR | NSP-SO | NSP-SR | NSP-SO |
| 1-layer | 1.03 | 0.92 | 2.99 | 3.08 |
| 2-layer | 0.86 | 0.71 | 2.66 | 2.80 |
| 3-layer | 0.78 | 0.63 | 2.80 | 2.91 |
| 4-layer | 0.79 | 0.68 | 2.50 | 2.63 |
| 5-layer | 0.78 | 0.63 | 2.75 | 2.87 |

## List of Figures

Figure 1. Surface energy (in eV) versus the number of Pu layers. The squares indicate NSP-SR values and the circles indicate NSP-SO values.

Figure 2. Work function (in eV) versus the number of Pu layers. The squares indicate NSP-SR values and the circles indicate NSP-SO values.

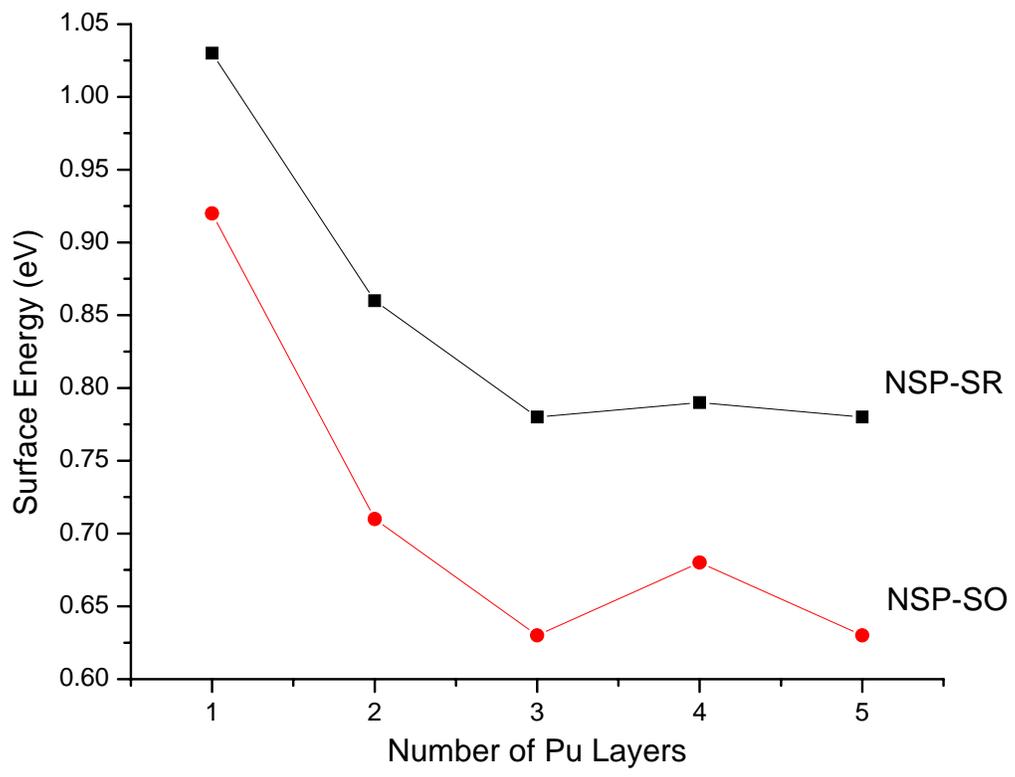

Fig.1 Surface energy (eV) vs. the number of Pu layers

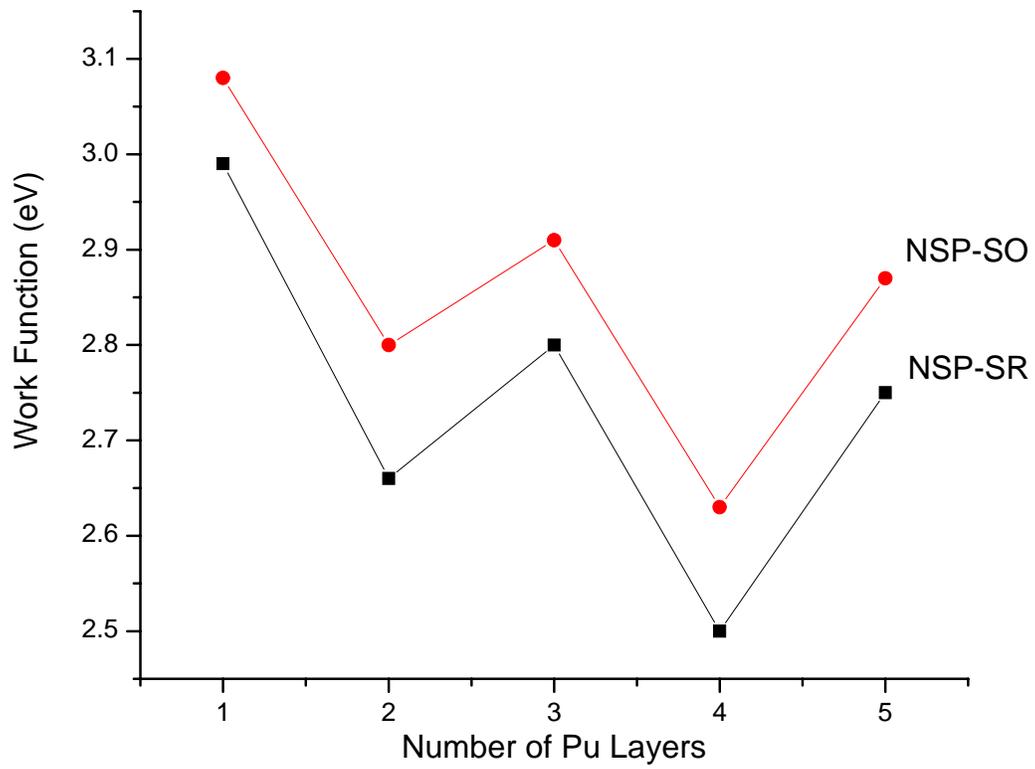

Fig. 2 Work function (eV) vs. the number of Pu layers.